\documentclass[a4paper,10pt]{article}
\usepackage{amsfonts, epsfig, amsmath, amssymb, color}

\newtheorem{theorem}{Theorem}[section]

\newcommand{\be}{\begin{equation}}
\newcommand{\ee}{\end{equation}}
\newcommand{\ben}{\begin{equation*}}
\newcommand{\een}{\end{equation*}}

\newcommand{\I}[1]{\mathbb{I}\{#1\} }

\begin{document}
\title{Cooling Down L\'evy Flights}

\author{Ilya Pavlyukevich\footnote{Correspondence should be addressed to pavljuke@mathematik.hu-berlin.de}\\
Institut f\"ur Mathematik\\ Humboldt-Universit\"at zu Berlin\\
 Rudower Chaussee 25\\ 12489 Berlin Germany
}

\date{7 November 2006}

\maketitle

\begin{abstract}
Let $L(t)$ be a L\'evy flights process with a stability index 
$\alpha\in(0,2)$, and $U$ be an external multi-well potential. 
A jump-diffusion $Z$ satisfying a stochastic differential equation 
$dZ(t)=-U'(Z(t-))dt+\sigma(t)dL(t)$
describes an evolution of a L\'evy particle of an `instant temperature' 
$\sigma(t)$ in an external force field. The temperature is supposed to decrease 
polynomially fast, i.e.\ $\sigma(t)\approx t^{-\theta}$ for some 
$\theta>0$. 
We discover two different cooling regimes. 
If $\theta<1/\alpha$ (slow cooling), 
the jump diffusion $Z(t)$ has a non-trivial limiting distribution as $t\to \infty$,
which is concentrated at the potential's local minima. 
If $\theta>1/\alpha$ (fast cooling)
the L\'evy particle gets trapped in one of the potential wells.
\end{abstract}

PACs numbers: 05.40.Fb, 02.50.Ey, 02.50.Fz, 02.50.Ga

\numberwithin{equation}{section}

\section{Introduction}

In this paper we study a L\'evy flights dynamics in an external multi-well 
potential in the annealed regime.
We are motivated by the problem of random search of the global 
minimum of an unknown function
$U$ with help of simulated annealing. For simplicity, we consider a one-dimensional case. 
Let $U$ be a multi-well potential satisfying some regularity conditions.  
Classical continuous time simulated annealing consists in an examination of a time non-homogeneous Smoluchowski
diffusion
\begin{equation}
\label{eq:1}
d\hat{Z}(t)=-U'(\hat{Z}(t))dt+\hat{\sigma}(t)dW(t)
\end{equation}
with a temperature $\hat{\sigma}(t)\to 0$ as $t\to+\infty$. For small values of $\hat{\sigma}(t)$,
the process $\hat{Z}$ spends most of the time in small neighbourhoods of the 
potential's local
minima and makes occasional transitions between the adjacent wells. It is possible to choose
an appropriate cooling schedule $\hat{\sigma}(t)$, such that the diffusion 
settles down near the
global maximum of $U$. Indeed, one should take 
$\hat{\sigma}^2(t)\approx \frac{\theta}{\ln(\lambda+t)}$, the parameter $\theta>0$ being a
cooling rate and $\lambda>1$ parameterising the initial temperature. Then there is a critical
value $\hat{\theta}>0$, such that $\hat{Z}(t)$ converges in probability to the global
minimum of $U$ if $\theta>\hat{\theta}$, and the convergence fails if $0<\theta<\hat{\theta}$. 
Moreover, the critical value $\hat{\theta}$ is a logarithmic growth rate of the principal
non-zero eigenvalue $\lambda^1(\sigma)$ of the generator of the time-homogeneous diffusion
\begin{equation}
d\hat{X}(t)=-U'(\hat{X}(t))dt+\sigma dW(t),
\end{equation} 
i.e.\ $|\lambda^1(\sigma)|\propto \exp(-\hat{\theta}/\sigma^2)$.
Heuristic justification for the convergence is as follows. 
The principal non-zero eigenvalue $\lambda^1(\sigma)$
determines the convergence rate of $\hat{X}$ to its invariant measure 
$\mu_\sigma(dx)=c_\sigma\exp(-2U(x)/\sigma^2)dx$, $c_\sigma$ being a normalising factor.
Thus, for any continuous positive function $f$ we have an estimate
\begin{equation}
|\mathbf{E}_x f(X(t)) - \int f(x)\mu_\sigma(dx)|
\leq C e^{-|\lambda^1(\sigma)|t}.
\end{equation} 
The weak limit of the invariant measures $\mu_\sigma(dy)$ as $\sigma\to 0$ 
is a Dirac mass at the potential's global minimum. 
For small values of $\sigma(t)$, the dynamics of $\hat{Z}$ reminds of a 
dynamics of $\hat{X}$. 
Thus  $\hat{Z}(t)$ has enough time to settle down in the deepest potential well
if $\sigma(t)$ is such that
\begin{equation}
t|\lambda^1(\sigma(t))|\to \infty \quad 
\Leftrightarrow\quad 
\frac{t}{(\lambda+t)^{\hat{\theta}/\theta}}\to \infty
\quad 
\Leftrightarrow\quad 
\theta>\hat{\theta},\quad t\to+\infty.
\end{equation}
The logarithmic decrease rate of $\sigma$ was obtained in the seminal paper
\cite{GemanG-84}.
Further mathematical results on classical continuous time simulated 
annealing can be found in \cite{ChiangHS-87,HwangS-90,HwangS-92,HolleyS-88,HolleyKS-89}. 
We also refer the reader to the review paper \cite{Gidas-95} and further references 
therein.

Our research is motivated by the paper \cite{SzuH-87} by Szu and Hartley, where they
introduced the 
so-called \textit{fast simulated annealing}
which allows to perform a non-local search of the deepest well.
Fast simulated annealing process in the sense of \cite{SzuH-87} is a discrete time 
Markov chain, where the states are
obtained from the Euler approximation of \eqref{eq:1} driven not by Gaussian noise but 
\textit{Cauchy} noise. 
The new state is accepted according to the Metropolis algorithm 
(see \cite{MetropolisRRT-53}) with Boltzmann acceptance probability which equals $1$, 
if the potential value in this state is smaller, i.e.\ the new position is `lower'
in the potential landscape. If the new position is `higher', it is accepted with the probability 
$\sim \exp(-\Delta U/\sigma)$, where $\Delta U$ is the difference of the potential values in the 
new and the old states, and $\sigma$ is a decreasing `temperature' parameter. The 
advantage of this method consists in faster transitions between the potential wells due to 
the 
heavy tails of Cauchy distribution. Moreover, the authors claim that the optimal cooling
rate is algebraic, $\sigma(t)\approx t^{-1}$, which also accelerates convergence.

In this paper, we consider a continuous-time 
L\'evy flights counterpart of the diffusion \eqref{eq:1}. 
Our goal is to study the asymptotic properties of the system in dependence of a
cooling schedule. We notify the reader, that in regimes where a L\'evy flights
process converges to some limiting distribution, it does not locate the global minimum
of $U$, but reveals the spatial structure of the potential. 

This paper contains a heuristic derivation of results, 
which are proved rigorously in \cite{Pavlyukevich-06a}. 
It can be seen as a sequel of \cite{ImkellerP-06,ImkellerP-06a,ImkellerP-06b} where a 
small-noise dynamics of L\'evy flights in external potentials were studied. We emphasise 
that our methods are purely probabilistic.

\section{Object of study and results}

\subsection{L\'evy flights}

Let $L=(L(t))_{t\geq 0}$ be a L\'evy flights (LF) process of index $\alpha\in(0,2)$, 
i.e.\ a non-Gaussian stable symmetric L\'evy
process with marginals having the Fourier transform
\begin{equation}
\label{eq:f}
\mathbf{E} e^{i\omega L(t)}=e^{-c(\alpha)t|\omega|^\alpha},
\quad c(\alpha)=2\int_0^\infty\frac{1-\cos{y}}{y^{1+\alpha}}\,dy=2|\cos(\frac{\pi\alpha}{2})\Gamma(-\alpha)|.
\end{equation}
In our analysis we shall use the L\'evy-Khinchin representation of the characteristic function of $L(t)$, namely
\begin{equation}
\label{eq:lh}
\mathbf E e^{i\omega L(t)}=
\exp\left\lbrace t
\int_{\mathbb R\backslash\{0\}} \left[ e^{i\omega y}-1
-i\omega y \I{|y|\leq 1}\right]
\frac{dy}{|y|^{1+\alpha}}\right\rbrace,
\end{equation}
where $\I{A}$ denotes the indicator function of a set $A$.
The most important ingredient of the representation
\eqref{eq:lh} is the so called \textit{L\'evy (jump) measure} of the random process
$L$ given by
\begin{equation}
\nu(A)=\int_{A\backslash\{0\}}\frac{dy}{|y|^{1+\alpha}}, \quad
A\,\,\mbox{ is a Borel set in}\,\,\mathbb{R}.
\end{equation}
Note, that some authors prefer another parametrisation of LFs with the Fourier transform $\exp(-t|\omega|^\alpha)$.
In this paper we use the representation \eqref{eq:f}
due to a simpler form of the L\'evy measure. 

The measure $\nu$ controls the intensity and sizes of the jumps of
the L\'evy flights process. Let $\Delta L(t)=L(t)-L(t-)$ be the random jump 
size of $L$ at a time instance $t$, $t>0$, and the number of jumps belonging to the set $A$ on the time interval $(0,t]$ be denoted 
by $N(t, A)$, i.e.\
\begin{equation}
N(t,A)=\sharp\{s: (s,\Delta L_s)\in (0,t]\times A\}.
\end{equation}
Then the random variable $N(t, A)$ has a Poisson distribution with mean
$t\nu(A)$ (which can possibly be infinite). 
Note, that for any stability index $\alpha\in(0,2)$, the L\'evy measure of 
any neighbourhood of $0$ is infinite, hence LFs make infinitely many very 
small jumps on any time interval.
The tails of the density $|y|^{-1-\alpha}dy$   
determine big jumps of LFs. Thus, $\mathbf{E}|L(t)|^\delta<\infty$, $t>0$, 
iff $\int_{|y|\geq 1}|y|^\delta\nu(dy)<\infty$
iff $\delta<\alpha$.

\subsection{External potential}

We assume that the external potential $U$ is smooth and has $n$ local minima $m_i$ and $n-1$ local maxima $s_i$ enumerated in the increasing order, i.e.\
\begin{equation}
-\infty=s_0<m_1<s_1<\cdots < s_{n-1}<m_n<s_{n}=+\infty.
\end{equation}
We assume also that local extrema are non-degenerate, i.e.\ $U''(m_i)>0$ and $U''(s_i)<0$, and the potential 
increases fast at infinity, i.e.\
$|U'(x)|>|x|^{1+c}$, $|x|\to\infty$ for some $c>0$.

Under the assumptions on $U$, the deterministic dynamical system
\begin{equation}
\label{eq:x0}
X^0_x(t)=x-\int_0^t U'(X^0_x(u))\, du
\end{equation}
has $n$ domains of attraction $\Omega_i=(s_{i-1},s_i)$ with asymptotically stable 
attractors $m_i$. We note that if $x\in \Omega_i$ then $X^0_x(t)\in \Omega_i$ for all $t\geq 0$,
i.e.\ the deterministic trajectory cannot pass between different domains of attraction.
Denote $B_i=\{x:|m_i-x|\leq \Delta\}$ a $\Delta$-neighbourhood of the attractor $m_i$.
We suppose that $\Delta$ is small enough, so that $B_i\subset \Omega_i$, $1\leq i\leq n$.
Due to the rapid increase of $U'$ at infinity, the return of $X^0_x(t)$ from $\pm\infty$
to $B_1$ or $B_n$ occurs in finite time.

\subsection{Small constant temperature}

First, we consider L\'evy flights $L$ in the potential $U$ in the regime of 
small constant temperature.
The resulting random dynamics is described by the stochastic differential equation
\begin{eqnarray}
X^\varepsilon_x(t)&=x-\int_0^t U'(X^\varepsilon_x(u-))\, du+\varepsilon L(t), \quad x\in\mathbb{R},\,\,t\geq 0.
\end{eqnarray} 
The properties of $X^\varepsilon$ are studied in our previous 
paper \cite{ImkellerP-06a} in the case of a double-well potential. 
The general multi-well case is studied in \cite{ImkellerP-06b}. Below we formulate 
the results.

For any $\Delta>0$ sufficiently small, in the limit $\varepsilon\to 0$, the process $X^\varepsilon$ spends an overwhelming 
proportion of time in the set $\cup_{i=1}^n B_i$ making occasional abrupt jumps between 
different neighbourhoods $B_i$.
Thus, the knowledge of the transition times and probabilities is essential
for understanding the asymptotic properties of $X^\varepsilon$.
Let $T^i_x(\varepsilon)=\inf\{t\geq 0: X^\varepsilon_x(t)\in \cup_{j\neq i}B_j\}$. For $x\in B_i$, the stopping time
$T^i_x(\varepsilon)$ denotes the first transition time to a $\Delta$-neighbourhood of 
a minimum of a different well.
Then we have the following result.
\begin{theorem}[constant temperature, transitions]
\label{th:T}
For $x\in B_i$, $1\leq i\leq n$, the following estimates hold in the limit $\varepsilon\to 0$:
\begin{align}
\label{eq:t1}
&\mathbf{P}_x\left( X^\varepsilon(T^i(\varepsilon))\in B_j\right) \to\frac{q_{ij}}{q_i}, \quad i\neq j,\\
\label{eq:t2}
&\varepsilon^\alpha T^i_x(\varepsilon)\stackrel{d}{\to} \exp(q_i),\\
\label{eq:t3}
&  \varepsilon^\alpha \mathbf{E}_x T^i(\varepsilon)\to \frac{1}{q_i},
\end{align}
where
\begin{align}
\label{eq:q}
q_{ij}&
=\frac{1}{\alpha}
\left|\frac{1}{|s_{j-1}-m_i|^\alpha}-\frac{1}{|s_j-m_i|^\alpha}\right|,\quad i\neq j,\\
q_i&=\sum_{j\neq i}q_{ij}=\frac{1}{\alpha}\left( \frac{1}{|s_{i-1}-m_i|^\alpha}
+\frac{1}{|s_i-m_i|^\alpha}\right),
\end{align}
and ``$\stackrel{d}{\to}$'' denotes convergence in distribution.
\end{theorem}

As we see, the transition times between the wells of $X^\varepsilon$ are asymptotically 
exponentially distributed in the limit of small noise, and hence \textit{unpredictable}, 
due to the memoryless property of the exponential law.
The transition probabilities between the wells are noise independent and strictly positive.
Thus, $X^\varepsilon$ reminds of a Markov process on a finite state space. Indeed, the following theorem 
holds. 
\begin{theorem}[constant temperature, metastability]
\label{th:meta}
If $x\in \Omega_i$,
$1\leq i\leq n$, then for $t>0$
\begin{eqnarray}
X^\varepsilon_x\left( \frac{t}{\varepsilon^\alpha}\right) \to Y_{m_i}(t),\quad \varepsilon\to 0,
\end{eqnarray}
in the sense of finite-dimensional distributions, where $Y=(Y_y(t))_{t\geq 0}$ is a Markov process on a state space $\{m_1,\dots, m_n\}$ with the infinitesimal generator $Q=(q_{ij})_{i,j=1}^n$,
$q_{ij}$ being defined in \eqref{eq:q}, $q_{ii}=-q_i$.
\end{theorem}

Since none of the entries $q_{ij}$ vanishes, the limiting Markov process $Y$ has a unique invariant distribution 
$\pi=(\pi_1,\dots,\pi_n)^T$, which can be calculated from the matrix equation $Q^T\pi=0$.

\subsection{Decreasing temperature}

In the annealed regime, the dynamics of L\'evy flights is characterised by 
the time non-homogeneous equation
\begin{eqnarray}
\label{eq:Z}
Z^\lambda_{s,z}(t)&=z-\int_s^t U'(Z^\lambda_{s,z}(u-))\, du+\int_s^t \frac{dL(u)}{(\lambda+u)^\theta},
\quad z\in\mathbb{R},\,\,0\leq s\leq t,
\end{eqnarray} 
where a positive parameter $\theta$ denotes the \textit{cooling rate}, and $\lambda>0$ determines the initial
temperature, which equals to $(\lambda+s)^{-\theta}$.

It is easily seen from \eqref{eq:Z}, that the evolution of the process starting
at time $s\geq 0$ is the same as that of the process starting at time zero with a different initial temperature, namely
\begin{equation}
\label{eq:mp}
(Z^\lambda_{s,z}(s+t))_{t\geq 0}\stackrel{d}{=}(Z^{\lambda+s}_{0,z}(t))_{t\geq 0},
\end{equation}
and thus the particular values of $s$ or $\lambda$ do not influence asymptotic
properties of the 
process in the limit $t\to\infty$. However, since our theory will work
for low temperatures, it is often convenient to study
the dynamics not for large values of $s$ and $t$ but for large
values of $\lambda$.

The goal of this paper is to study the limiting behaviour of $Z^\lambda_{0,z}(t)$ as $t\to \infty$
in dependence of the cooling rate $\theta$, `initial temperature' $\lambda$ and initial
point $z$.

Similarly to the classical Gaussian case discussed in the introduction,
the candidate for the limiting law of $Z^\lambda_{0,z}(t)$ is the invariant 
distribution $\pi$ of the Markov chain $Y$ from Theorem~\ref{th:meta}.
Furthermore, we have to distinguish between two different cooling regimes. 

As in the previous section, for $1\leq i\leq n$, consider the stopping times
\begin{equation}
\tau_{s,z}^{i,\lambda}=\inf\{u\geq s:Z_{s,z}^\lambda(u)\in \cup_{j\neq i}B_j\}.
\end{equation}
If $z\in B_i$ then $\tau_{s,z}^{i,\lambda}$ denotes the \textit{transition time} from a
$\Delta$-neighbourhood of $m_i$ to a $\Delta$-neighbourhood of some other potential's minimum. 
For all $j\neq i$ we also consider the corresponding
\textit{transition probabilities} 
$\mathbf{P}_{s,z}(Z^\lambda(\tau^{i,\lambda})\in B_j)$. Then the following analogue of Theorem~\ref{th:T}
holds.
\begin{theorem}[slow cooling, transitions]
\label{th:tau}
Let $\theta<1/\alpha$. For $z\in B_i$, $1\leq i\leq n$, the following estimates hold in the
limit of small initial temperature, i.e. when $\lambda\to+\infty$:
\begin{align} 
\label{eq:p}
&\mathbf{P}_{0,z}(Z^\lambda(\tau^{i,\lambda})\in B_j)
\to\frac{q_{ij}}{q_i},\quad i\neq j,\\
\label{eq:tau}
&\frac{\mathbf{E}_{0,z} \tau^{i,\lambda}}{\lambda^{\alpha\theta}}\to \frac{1}{q_i},
\end{align} 
$q_i$ and $q_{ij}$ being defined in \eqref{eq:q}.
\end{theorem}

\begin{theorem}[slow cooling, convergence]
\label{th:slow}
Let $\theta<1/\alpha$. 
Then for any $\lambda>0$, $z\in\mathbb{R}$, the law of $Z_{0,z}^\lambda(t)$ converges weakly to
the measure $\pi$, i.e.\
for any continuous and bounded function $f$ we have
\begin{equation}
\mathbf{E}_{0,z}f(Z^\lambda(t))\to \sum_{j=1}^n f(m_j)\pi_j, \quad t\to\infty.
\end{equation}
\end{theorem}

If the cooling rate $\theta$ is above the threshold $1/\alpha$, the solution $Z^\lambda$ gets trapped in one of the wells and thus the convergence fails. Consider the first exit time 
from the $i$-th well
\begin{equation}
\sigma^{i,\lambda}_{s,z}=\inf\{t\geq 0: Z^\lambda_{s,z}\notin \Omega_i \}.
\end{equation}

Then the following trapping result holds. 
\begin{theorem}[fast cooling, trapping]
\label{th:fast}
Let $\theta>1/\alpha$.  For $z\in B_i$, $1\leq i\leq n$,  
\begin{equation}
\mathbf{P}_{0,z}(\sigma^{i,\lambda}<\infty)
=\mathcal{O}\left(\frac{1}{\lambda^{\alpha\theta-1}}\right),\quad\lambda\to\infty.
\end{equation}
Consequently, $\mathbf{E}_{0,z}\sigma^{i,\lambda}=\infty$.
\end{theorem}

In the subsequent section we sketch the proof of Theorems~\ref{th:tau}--\ref{th:fast} and discuss the results.

\section{Predominant behaviour of the annealed process}

Our study of the random process $Z^\lambda$ is based on probabilistic analysis of its
sample paths. We use the decomposition
of the process $L$ into small- and big-jump parts similar to that used in \cite{ImkellerP-06a}.
Thus we refer the reader to that paper for details, and sketch the idea briefly.

\subsection{Big and small jumps of a L\'evy flights process}

With help of the L\'evy-Khinchin formula \eqref{eq:lh}, we decompose the 
process $L$ into a sum of two independent L\'evy
processes with relatively small and big jumps.  For any cooling rate $\theta>0$, we
introduce two new L\'evy measures by setting
\begin{eqnarray}
\nu_\xi^\lambda(A) &= \nu\left(A\cap \{x:|x|\leq \lambda^{\theta/2}\}\right),\\
\nu_\eta^\lambda(A)& = \nu\left(A\cap \{x:|x|> \lambda^{\theta/2}\}\right),
\end{eqnarray}
and two L\'evy processes $\xi^\lambda$ and $\eta^\lambda$
with the corresponding Fourier transforms:
\begin{eqnarray}
\mathbf E e^{i\omega \xi^\lambda_t}&=
\exp\left\lbrace
t\int_{\mathbb R\backslash\{0\}} \left[e^{i\omega y}-1-i\omega y \I{|y|\leq 1}\right]
\nu_\xi^\lambda(dy)  \right\rbrace,\\
\mathbf E e^{i\omega \eta^\lambda_t}&=
\exp\left\lbrace
t\int_{\mathbb R\backslash\{0\}} \left[ e^{i\omega y}-1-i\omega y \I{|y|\leq 1}\right]
\nu_\eta^\lambda(dy)\right\rbrace.
\end{eqnarray}
It is clear that, the processes $\xi^\lambda$ and $\eta^\lambda$ are independent and
$L=\xi^\lambda+\eta^\lambda$.

Since $\nu^\lambda_\xi(\mathbb{R})=\infty$, the process
$\xi^\lambda$ makes infinitely many jumps on each time
interval. Its jumps  are, however, bounded by the threshold
$\lambda^{\theta/2}$, i.e.\ $|\Delta\xi_t^\lambda|\leq \lambda^{\theta/2}$. 
Thus $\xi^\lambda_t$ has a finite variance, and more generally moments of all
orders.

On the contrary, the L\'evy measure of the process
$\eta^\lambda$ is finite, and its mass equals
\begin{equation}
\beta_\lambda=\nu_\eta^\lambda(\mathbb{R})
=\int_{-\infty}^{-\lambda^{\theta/2}} \frac{dy}{|y|^{1+\alpha}}
+\int_{\lambda^{\theta/2}}^\infty \frac{dy}{y^{1+\alpha}}
=2\int_{\lambda^{\theta/2}}^\infty \frac{dy}{y^{1+\alpha}}
=\frac{2}{\alpha}\lambda^{-\alpha\theta/2}.
\end{equation}
Hence, $\eta^\lambda$ is a compound Poisson process with jumps
of absolute value larger than $\lambda^{\theta/2}$. Let $\tau^\lambda_k$ and
$W^\lambda_k$, $k\geq 0$, be the jump arrival times and jump sizes of $\eta^\lambda$ 
under the
convention $\tau^\lambda_0=W^\lambda_0=0$. Then the inter-arrival times
$T^\lambda_k=\tau^\lambda_k-\tau^\lambda_{k-1}$, $k\geq 1$, are independent and
exponentially distributed with mean $\beta_\lambda^{-1}$. The jump sizes $W^\lambda_k$
are also independent random variables with
the probability distribution function given by
\begin{equation}
\label{eq:w}
\mathbf{P}(W^\lambda_k< u)
=\frac{\nu_\eta^\lambda(-\infty,u)}{\nu_\eta^\lambda(\mathbb{R})}
=\frac{1}{\beta_\lambda}\int_{-\infty}^u \I{|y|> \lambda^{\theta/2}}
\nu^\lambda_\eta(dy).
\end{equation}
Finally, we can represent the random perturbation in \eqref{eq:Z} as a sum of two processes, namely,
\begin{equation}
\int_0^t\frac{dL(u)}{(\lambda+u)^\theta}
=\int_0^t\frac{d\xi^\lambda_u}{(\lambda+u)^\theta}
+\sum_{k=1}^\infty \frac{W^\lambda_k}{(\lambda+\tau^\lambda_k)^\theta}
\I{t\geq \tau_k}.
\end{equation}

\subsection{Predominant behaviour}

Consider now the process $Z^\lambda_{0,z}$ given by equation
\eqref{eq:Z}. On the inter-arrival intervals $[\tau^\lambda_{k-1}, \tau^\lambda_k)$,
$k\geq 1$, it is driven only by the process
$\varphi_t^\lambda=\int_0^t (\lambda+u)^{-\theta}d\xi^\lambda_u$, 
and at the time instants $\tau^\lambda_k$ it
makes jumps of the size $W^\lambda_k/(\lambda+\tau_\lambda^k)^\theta$. 
Recall that the jumps of
$\xi^\lambda$ are bounded by $\lambda^{\theta/2}$, hence the jumps sizes 
of $\varphi^\lambda$
tend to zero as $\lambda\to\infty$ for all $t\geq 0$, i.e.\ 
\begin{equation}
|\Delta \varphi^\lambda_t|\leq \frac{\lambda^{\theta/2}}{(\lambda+t)^\theta}\leq \frac{1}{\lambda^{\theta/2}}.
\end{equation}
The variance of $\varphi_t^\lambda$ tends to zero in the
limit of large $\lambda$, and the random trajectory $Z^\lambda_{0,z}(t)$
can be seen as a small random perturbation of the deterministic
trajectory $X^0_z(t)$ of the underlying dynamical system on the
intervals $[\tau^\lambda_{k-1},\tau^\lambda_k)$. 
Consider a well $\Omega_i$ with a minimum $m_i$. Let initial points $z$ be
away from the unstable points $s_{i-1}$ and $s_i$, namely 
$z\in(s_{i-1}+\lambda^{-\gamma}, s_i-\lambda^{-\gamma})$ for some positive $\gamma$.
Then the deterministic trajectory $X_z^0(t)$ reaches a $\lambda^{-\gamma}$-neighbourhood of $m_i$ in at most logarithmic time $\mathcal{O}(\ln\lambda)$.
Since the periods between the big jumps are essentially longer, i.e.\
\begin{equation}
\mathbf{E} T_k^\lambda=\mathbf{E}(\tau_{k}^\lambda-\tau_{k-1}^\lambda)
= \frac{\alpha}{2}\lambda^{\alpha\theta/2} \gg \mathcal{O}(\ln\lambda),
\end{equation} 
we can show that with probability close to $1$, the random trajectory $Z^\lambda$ is located 
in a small neighbourhood of $m_i$ before the big jump. 

Thus we can summarise the pathwise behaviour of $Z^\lambda_{0,z}$ for large values of
$\lambda$ as follows: 
\begin{equation}
\begin{aligned}
\label{eq:t}
&Z^\lambda_{0,z}(0)=z\in (s_{i-1}+\lambda^{-\gamma}, s_i-\lambda^{-\gamma}),\\
&Z^\lambda_{0,z}(\tau^1_\lambda-)\approx m_i,\\
&Z^\lambda_{0,z}(\tau^1_\lambda)\approx
m_i+\frac{W_1^\lambda}{(\lambda+\tau_1^\lambda)^\theta}\in (s_{i-1}+\lambda^{-\gamma}, s_i-\lambda^{-\gamma})\\
&Z^\lambda_{0,z}(\tau^2_\lambda-)\approx m_i,\\
&\cdots\\
&Z^\lambda_{0,z}(\tau^k_\lambda-)\approx m_i,\\
&Z^\lambda_{0,z}(\tau^k_\lambda)\approx
m_i+\frac{W_k^\lambda}{(\lambda+\tau_k^\lambda)^\theta}\in (s_{j-1}+\lambda^{-\gamma}, s_j-\lambda^{-\gamma}), 
\,\,j\neq i,\\
&Z^\lambda_{0,z}(\tau^{k+1}_\lambda-)\approx m_j,\\
&\cdots,
\end{aligned}
\end{equation}
whereas on the intervals $[\tau^\lambda_{k-1},\tau^\lambda_k)$ the process $Z^\lambda$ follows the
deterministic trajectory $X^0$.
Thus, since we know the initial location of the particle, 
as well as the jump sizes $W_k^\lambda$ and jump times $\tau_k^\lambda$,
we can catch the essential features of the random path $Z^\lambda$.

Of course, we have to be carefull when dealing with trajectories which occasionally enter
the $\lambda^{-\gamma}$-neighbourhoods of the saddle points $s_i$, where the force field
$U'$ becomes insignificant. In these neighbourhoods, the L\'evy particle has no strong 
deterministic drift which brings it to a certain well's minimum. Thus we cannot decide
whether $Z^\lambda$ converges to $m_i$ or to $m_{i+1}$. However, in the limit
$\lambda\to\infty$, the probability that $Z^\lambda$ jumps from a 
neighbourhood of $m_i$ to a 
$\lambda^{-\gamma}$-neighbourhood of some $s_j$ is negligible. In our further 
exposition, we do not consider the unstable dynamics in these
$\lambda^{-\gamma}$-neighbourhoods 
and assume that \eqref{eq:t} holds for all $z\in\Omega_i$.
Interested readers can find rigorous arguments in \cite{Pavlyukevich-06a}.

\section{Transitions between the wells in the slow cooling regime\label{s:kl}}

In this section we justify the limits \eqref{eq:p} and \eqref{eq:tau} from 
Theorem~\ref{th:tau}.

First, we obtain the mean value of the first exit time $\sigma^{i,\lambda}$
form the well $\Omega^i$ in the limit $\lambda\to\infty$.
Indeed, $Z^\lambda$ can roughly leave $\Omega_i$ only
at one of the time instants $\tau^\lambda_k$ when 
$m_i+W^\lambda_k/(\lambda+\tau^\lambda_k)^\theta\notin \Omega_i$.
We can therefore calculate the mean value of $\sigma^{i,\lambda}$
using the full probability formula:
\begin{equation}
\label{eq:m}
\begin{aligned}
&\mathbf{E}_{0,z}\sigma^{i,\lambda}
\approx \sum_{k=1}^{\infty}
\mathbf{E}\left[\tau^\lambda_k \I{ \sigma^{i,\lambda}=\tau^\lambda_k} \right]  \\
&\approx\sum_{k=1}^{\infty}
\mathbf{E}\left[\tau^\lambda_k\cdot
\I{m_i+ \frac{W^\lambda_1}{(\lambda+\tau_1^\lambda)^\theta}\in \Omega_i,\dots, 
m_i+ \frac{W^\lambda_{k-1}}{(\lambda+\tau_{k-1}^\lambda)^\theta}\in \Omega_i,
m_i+ \frac{W^\lambda_k}{(\lambda+\tau_k^\lambda)^\theta}\notin \Omega_i
}\right].
\end{aligned}
\end{equation}
Since the arrival times $\tau_1^\lambda,\tau_2^\lambda,\dots, \tau_k^\lambda$, are dependent, no straightforward
calculation of the 
expectations in the latter sum seems possible. However, we can estimate these expectations
from above and below.
Our argument is based on the inequalities  
$0<\tau_1^\lambda< \tau_2^\lambda<\cdots<\tau_k^\lambda$, and the obvious inclusions
\begin{equation}
\label{eq:inc}
\left\lbrace  m_i+ \frac{W^\lambda_{j}}{\lambda^\theta}\in \Omega_i\right\rbrace 
\subseteq
\left\lbrace  m_i+ \frac{W^\lambda_{j}}{(\lambda+\tau_{j}^\lambda)^\theta}\in \Omega_i\right\rbrace 
\subseteq
\left\lbrace  m_i+ \frac{W^\lambda_{j}}{(\lambda+\tau_{k}^\lambda)^\theta}\in \Omega_i\right\rbrace,
\,\,
1\leq j\leq k-1,
\end{equation}
where the probability of these events can be
calculated explicitly from \eqref{eq:w}, to yield the formula
\begin{equation}
\quad\mathbf{P}\left( m_i+\frac{W^\lambda_1}{(\lambda+t)^\theta} 
\notin \Omega_i\right)
=\frac{1}{\beta_\lambda}
\left( \int_{-\infty}^{-|m_i-s_{i-1}|(\lambda+t)^\theta} 
+\int_{(s_i-m_i)(\lambda+t)^\theta}^\infty \right)     \frac{dy}{|y|^{1+\alpha}}
=\frac{q_i}{\beta_\lambda(\lambda+t)^{\alpha\theta}}.
\end{equation}

\subsection{Mean transition time}

Let us obtain an estimate from above. Note that for each $k\geq 1$, the arrival time 
$\tau^\lambda_k$ is a sum of $k$ independent
exponentially distributed random variables $T^\lambda_j$ and thus has a    $\mbox{Gamma}(k,\beta_\lambda)$ distribution
with a probability density
$\beta_\lambda e^{-\beta_\lambda t}(\beta_\lambda t)^{k-1}/(k-1)!$, $t\geq 0$.
Then, applying the second inclusion in \eqref{eq:inc} we obtain
\begin{equation}
\begin{aligned}
&\mathbf{E}\left[\tau^\lambda_k\cdot
\I{m_i+ \frac{W^\lambda_1}{(\lambda+\tau_1^\lambda)^\theta}\in \Omega_i,\dots, 
m_i+ \frac{W^\lambda_{k-1}}{(\lambda+\tau_{k-1}^\lambda)^\theta}\in \Omega_i,
m_i+ \frac{W^\lambda_k}{(\lambda+\tau_k^\lambda)^\theta}\notin \Omega_i
}\right]\\
&\leq \mathbf{E}\left[\tau^\lambda_k\cdot
\I{m_i+ \frac{W^\lambda_1}{(\lambda+\tau_k^\lambda)^\theta}\in \Omega_i,\dots, 
m_i+ \frac{W^\lambda_{k-1}}{(\lambda+\tau_{k}^\lambda)^\theta}\in \Omega_i,
m_i+ \frac{W^\lambda_k}{(\lambda+\tau_k^\lambda)^\theta}\notin \Omega_i
}\right]\\
&=\int_0^\infty \beta_\lambda t e^{-\beta_\lambda t}\frac{(\beta_\lambda t)^{k-1}}{(k-1)!}
\mathbf{P}\left( m_i+ \frac{W^\lambda_1}{(\lambda+t)^\theta}
\in \Omega_i\right)^{k-1} 
\mathbf{P}\left( m_i+ \frac{W^\lambda_1}{(\lambda+t)^\theta}
\notin \Omega_i\right) dt\\
&=\int_0^\infty \beta_\lambda t e^{-\beta_\lambda t}\frac{(\beta_\lambda t)^{k-1}}{(k-1)!}
\left[1-\frac{q_i}{\beta_\lambda(\lambda+t)^{\alpha\theta}} \right]^{k-1} 
\frac{q_i}{\beta_\lambda(\lambda+t)^{\alpha\theta}} dt.
\end{aligned}
\end{equation}
Summation over $k$ yields
\begin{equation}
\begin{aligned}
\mathbf{E}_{0,z}\sigma^{i,\lambda}
&\lesssim
\int_0^\infty \beta_\lambda t e^{-\beta_\lambda t} \frac{q_i}{\beta_\lambda(\lambda+t)^{\alpha\theta}} \sum_{k=1}^{\infty}   
\frac{(\beta_\lambda t)^{k-1}}{(k-1)!}
\left[1-\frac{q_i}{\beta_\lambda(\lambda+t)^{\alpha\theta}} \right]^{k-1} dt\\
&=\int_0^\infty    \frac{q_i t }{(\lambda+t)^{\alpha\theta}} 
\exp\left( -\frac{q_i t }{(\lambda+t)^{\alpha\theta}} \right) dt,
\end{aligned}
\end{equation}
where `$\lesssim$' denotes an inequality up to negligible error terms.
Since $\alpha\theta<1$, the latter integral converges for all $\lambda>0$, and it is 
possible to evaluate it in the limit $\lambda\to+\infty$. 
Introducing a new variable $u=\frac{\lambda+t}{\lambda}$, we transform it
 to a Laplace type integral with big parameter, 
which can be evaluated asymptotically (see \cite[Chapter 3]{Olver-74}), i.e.\
\begin{equation}
\begin{aligned}
\mathbf{E}_{0,z}\sigma^{i,\lambda}
&\lesssim 
\lambda^{2-\alpha\theta}\int_1^\infty \frac{q_i(u-1)}{u^{\alpha\theta}}
\exp\left( -\frac{q_i (u-1)\lambda^{1-\alpha\theta} }{u^{\alpha\theta}} \right) du\\
&\approx \lambda^{2-\alpha\theta}\int_1^\infty q_i(u-1)
\exp\left( -q_i (u-1)\lambda^{1-\alpha\theta}  \right) du\\
&=q_i^{-1}\lambda^{\alpha\theta}.
\end{aligned}
\end{equation}
Applying analogously the first inclusion from \eqref{eq:inc} to the first $k-1$ jumps, 
we obtain the estimate from below:
\begin{equation}
\begin{aligned}
&\mathbf{E}_{0,z}\sigma^{i,\lambda}
\gtrsim \sum_{k=1}^{\infty}
\mathbf{E}\left[\tau^\lambda_k\cdot
\I{m_i+ \frac{W^\lambda_1}{\lambda^\theta}\in \Omega_i,\dots, 
m_i+ \frac{W^\lambda_{k-1}}{\lambda^\theta}\in \Omega_i,
m_i+ \frac{W^\lambda_k}{(\lambda+\tau_k^\lambda)^\theta}\notin \Omega_i
}\right]\\
&\geq \sum_{k=1}^{\infty}\int_0^\infty t\beta_\lambda e^{-\beta_\lambda t}
\frac{(\beta_\lambda t)^{k-1}}{(k-1)!}
\left[1-\frac{q_i}{\beta_\lambda \lambda^{\alpha\theta}} \right]^{k-1} 
\frac{q_i}{\beta_\lambda(\lambda+t)^{\alpha\theta}} dt\\
&=\int_0^\infty \frac{q_i t }{(\lambda+t)^{\alpha\theta}} 
\exp\left( -\frac{q_i t }{\lambda^{\alpha\theta}} \right) dt
= \lambda^{2-\alpha\theta} \int_1^\infty \frac{q_i(u-1)}{u^{\alpha\theta}}
\exp\left( -q_i(u-1)\lambda^{1-\alpha\theta} \right)\\
&\approx \lambda^{2-\alpha\theta}\int_1^\infty q_i(u-1)
\exp\left( -q_i (u-1)\lambda^{1-\alpha\theta}  \right) du
=q_i^{-1}\lambda^{\alpha\theta} .
\end{aligned}
\end{equation}
Surprisingly, the estimates from below and above coincide, and thus give the asymptotic value
of the mean life time of the slowly cooled L\'evy particle in a potential well.

To obtain the limit \eqref{eq:tau} for the mean transition time $\tau^{i,\lambda}$
between the sets $B_i$ and $\cup_{j\neq i}B_j$,
we note that
at the exit time $\sigma^{i,\lambda}$ the process $Z^\lambda$ enters some of the
wells $\Omega_j$, $j\neq i$, with high probability follows the deterministic trajectory, and reaches a $\Delta$-neighbourhood
of a well's minimum in a time of the order $\mathcal{O}(\ln\lambda)$, 
which is negligible in comparison with
$\lambda^{\alpha\theta}$. Thus the limit \eqref{eq:tau} holds.

\subsection{Transition probability}

To calculate the transition probability between the wells, it suffices to obtain an estimate from
below. Similarly to the estimate of the mean exit time, we have
\begin{equation}
\begin{aligned}
&\mathbf{P}_{0,z}(Z^\lambda(\sigma^{i,\lambda})\in\Omega_j)
\gtrsim \sum_{k=1}^{\infty}
\mathbf{P}\left(m_i+ \frac{W^\lambda_1}{\lambda^\theta}\in \Omega_i,\dots, 
m_i+ \frac{W^\lambda_{k-1}}{\lambda^\theta}\in \Omega_i,
m_i+ \frac{W^\lambda_k}{(\lambda+\tau_k^\lambda)^\theta}\in \Omega_j\right)\\
&\geq \sum_{k=1}^{\infty}\int_0^\infty \beta_\lambda e^{-\beta_\lambda t}
\frac{(\beta_\lambda t)^{k-1}}{(k-1)!}
\left[1-\frac{q_i}{\beta_\lambda \lambda^{\alpha\theta}} \right]^{k-1} 
\frac{q_{ij}}{\beta_\lambda(\lambda+t)^{\alpha\theta}} dt\\
&=\int_0^\infty \frac{q_{ij} }{(\lambda+t)^{\alpha\theta}} 
\exp\left( -\frac{q_i t }{\lambda^{\alpha\theta}} \right) dt
= \lambda^{1-\alpha\theta} \int_1^\infty \frac{q_{ij}}{u^{\alpha\theta}}
\exp\left( -q_{i}(u-1)\lambda^{1-\alpha\theta} \right)\\
&\approx \lambda^{1-\alpha\theta}\int_1^\infty q_{ij}
\exp\left( -q_i (u-1)\lambda^{1-\alpha\theta}  \right) du
=q_{ij}q_i^{-1}.
\end{aligned}
\end{equation}
With help of the equality $\sum_{j\neq i}q_{ij}q_i^{-1}=1$, we conclude that
$\mathbf{P}_{0,z}(Z^\lambda(\sigma^{i,\lambda})\in\Omega_j)\to q_{ij}q_i^{-1}$.
Finally, since after entering $\Omega_j$, the process $Z^\lambda$ reaches $B_j$ with high probability,
$\mathbf{P}_{0,z}(Z^\lambda(\sigma^{i,\lambda})\in\Omega_j)\approx
 \mathbf{P}_{0,z}(Z^\lambda(\tau^{i,\lambda})\in B_j)$, and \eqref{eq:p} holds.

\section{Convergence in the slow cooling regime}

\begin{figure}
\begin{center}
\includegraphics[width=.9\textwidth]{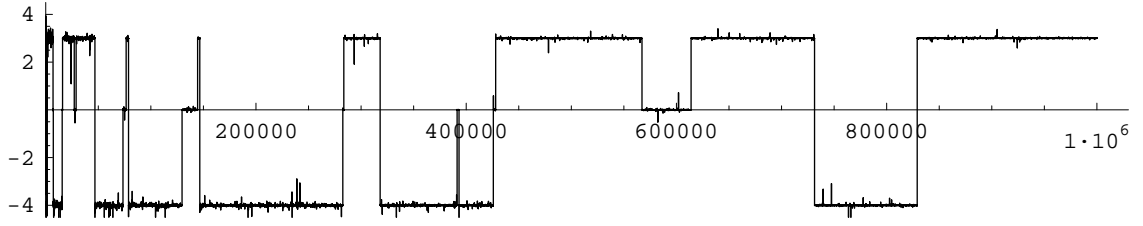}
\end{center}
\caption{A slow cooling of a L\'evy particle in a potential with 
local minima at $-4$, $0$ and $3$.\label{f:slow}}
\end{figure}

Figure~\ref{f:slow} illustrates the typical behaviour of $Z$ in the slow cooling 
regime, $\alpha\theta<1$. 
Roughly speaking, we can distinguish two different behaviours: chaotic and regular.

1.\ In general case, the initial temperature $\lambda^{-\theta}$ can be high, so that 
asymptotics of Theorem~\ref{th:tau} does not hold. 
Thus, the transitions of $Z^\lambda$ are chaotic until some time instance $T$, when 
the temperature $(\lambda+T)^{-\theta}$ becomes low enough, Theorem~\ref{th:tau} 
starts working.
Moreover, choosing the time $T$ sufficiently large, we
make the transition probabilities of $Z$ between the neighbourhoods $B_i$ 
close to $q_{ij}/q_i$ with any prescribed precision.
For brevity, we can also assume that 
$Z^\lambda_{0,z}(T)\in B_i$ for some $i$. 

2.\
Denote $\tau(k)$, $k\geq 0$, successive transition times after $T$ between 
different $B_j$, $1\leq j\leq n$, with $\tau(0)=T$ by convention. 
The mean values $\mathbf{E}_{0,z}\tau(k)$ are finite and can be calculated from Theorem~\ref{th:tau}. 
Indeed, if $\tau(k-1)=t_{k-1}$ and $Z^\lambda(\tau(k-1))=z_{k-1}\in B_i$, then
the conditional expectation of $\mathbf{E}_{0,z}\tau(k)$ equals
\begin{equation}
\begin{aligned}
&\mathbf{E}_{0,z}\left[\tau(k)|\tau(k-1)=t_{k-1},Z^\lambda(\tau(k-1))=z_{k-1}\in B_i\right]
=t_{k-1}+\mathbf{E}_{t_{k-1},z_{k-1}}\tau^{i,\lambda}\\
&=t_{k-1}+\mathbf{E}_{0,z_{k-1}}\tau^{i,\lambda+t_{k-1}}
\approx t_{k-1}+q_i^{-1}(\lambda+t_{k-1})^{\alpha\theta}<\infty.
\end{aligned}
\end{equation}
From the time instance $T$ on, $Z^\lambda$ makes transitions between 
the wells with probabilities close to $p_{ij}=q_{ij}/q_i$, $i\neq j$, where 
$p_{ii}=0$ by convention. The probabilities $p_{ij}$ determine a discrete time Markov chain $V(k)$ on 
$\{m_1,\dots, m_n\}$, such that $\mathbf{P}(V(k)=m_j|V(k-1)=m_i)=p_{ij}$. 
It is clear, that
$V$ has the unique invariant distribution $\pi$. Moreover, $V(k)$ converges to the
invariant distribution geometrically fast, i.e.\ there is $0<\rho<1$ such 
that for all $1\leq i,j\leq n$ and $k\geq 0$
\begin{equation}
\label{eq:V}
|\mathbf{P}_{m_i}(V(k)=m_j)-\pi_j|=\mathcal{O}({\rho^k}).
\end{equation}
With help of the asymptotic relation 
\begin{equation}
\mathbf{P}(Z^\lambda(\tau(k)\in B_j)|Z^\lambda(\tau(k-1)\in B_i))\approx p_{ij}=\mathbf{P}(V(k)=m_j|V(k-1)=m_i).
\end{equation}
one can show that the distributions of $Z(\tau(k))$ and $V(k)$ are also close
for $k\geq 1$, i.e.\ 
\begin{equation}
\mathbf{P}(Z^\lambda_{0,z}(\tau(k))\in B_j|Z^\lambda_{0,z}(T)\in B_i)\approx\mathbf{P}_{m_i}(V(k)=m_j).
\end{equation}
Hence, with help of \eqref{eq:V} for any prescribed accuracy level we can find $k_0\geq 1$ such that
for $k\geq k_0$ we have
\begin{equation}
\mathbf{P}_{0,z}(Z^\lambda(\tau(k))\in B_j)\approx\pi_j
\end{equation}
independently on the initial point $z$.

Finally, we note that
after time $T$, the process $Z^\lambda$ spends most of the time in the 
neighbourhoods $B_i$, and 
\begin{equation}
Z^\lambda_{0,z}(t)\approx Z^\lambda_{0,z}(\tau(k)),\quad \mbox{for }\, 
t\in [\tau(k),\tau(k+1)). 
\end{equation}
and thus if $t\geq \tau(k_0)$ then $\mathbf{P}_{0,z}(Z^\lambda(t)\in B_j)\approx \pi^0_j$ .

As we see, if $\alpha\theta<1$,
the process $Z^\lambda$ reminds of a peace-wise constant jump process on the state space
$\{m_1,\dots, m_n\}$. It
 never stops jumping between the wells of $U$, and the 
random sequence $Z^\lambda(\tau(k))$, $k\geq k_0$, behave as a stationary 
discrete time Markov chain with a distribution $\pi$.

\section{Trapping in the fast cooling regime}

The regime of fast cooling $\theta>1/\alpha$ is more simple.
We estimate the probability of the exit from a well. 
Since the exit occurs with high
probability only at the arrival times of the big jump process $\eta^\lambda$, we estimate
\begin{equation}
\begin{aligned}
&\mathbf{P}_{0,z}(\sigma^{i,\lambda}<\infty)\approx
\sum_{k=1}^\infty\mathbf{P}_{0,z}(\sigma^{i,\lambda}=\tau_k^\lambda)\\
&\lesssim\sum_{k=1}^\infty
\mathbf{P}\left(
m_i+ \frac{W^\lambda_1}{(\lambda+\tau_1^\lambda)^\theta}\in \Omega_i,\dots, 
m_i+ \frac{W^\lambda_{k-1}}{(\lambda+\tau_{k-1}^\lambda)^\theta}\in \Omega_i,
m_i+ \frac{W^\lambda_k}{(\lambda+\tau_k^\lambda)^\theta}\notin \Omega_i
\right)\\
&\leq\sum_{k=1}^\infty
\mathbf{P}\left(m_i+ \frac{W^\lambda_k}{(\lambda+\tau_k^\lambda)^\theta}\notin \Omega_i
\right)
=\int_0^\infty \beta_\lambda e^{-\beta_\lambda t} \frac{q_i}{\beta_\lambda(\lambda+t)^{\alpha\theta}}  \sum_{k=1}^\infty\frac{(\beta_\lambda t)^{k-1}}{(k-1)!} dt\\
&=\int_0^\infty    \frac{q_i }{(\lambda+t)^{\alpha\theta}} dt
=\frac{q_i}{\alpha\theta-1}\frac{1}{\lambda^{\alpha\theta-1}}\to 0, \quad \lambda\to\infty.
\end{aligned}
\end{equation} 
As a consequence we have infinite mean exit times $\mathbf{E}_{0,z}\sigma^{i,\lambda}=\infty$.
\begin{figure}
\begin{center}
\includegraphics[width=.9\textwidth]{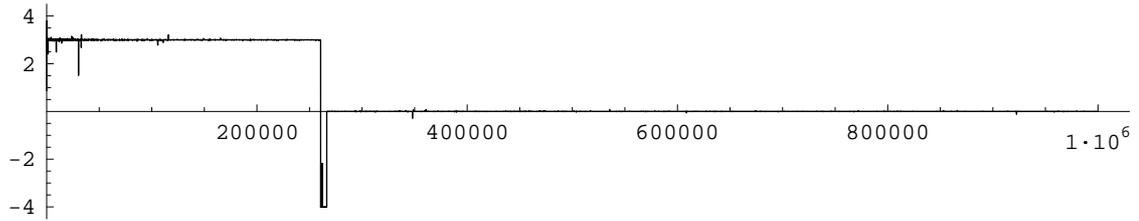}
\end{center}
\caption{A fast cooling of a L\'evy particle in a potential with 
local minima at $-4$, $0$ and $3$.\label{f:fast}}
\end{figure}
In other words, if $\theta>1/\alpha$, the dynamics of $Z^\lambda$ has two
qualitatively different regimes. First, for high temperatures, the
transitions between the wells are chaotic. Second, when the temperature is low enough,
the particle gets trapped in one of the wells, see Figure~\ref{f:fast}. 
In this case, there is no convergence
to the invariant measure $\pi$.

\section{Conclusion and discussion}

In this paper we studied the large time dynamics of a L\'evy particle 
in a multi-well external potential, with the temperature decreasing with time
as $1/t^\theta$. 

We discovered, 
that if the cooling is slow, i.e.\ $\theta<1/\alpha$, then the system reaches a
quasi-stationary regime where the transition probabilities between 
the wells converge to certain values which are explicitly
determined in terms of the potential's spatial geometry. 
Moreover, the mean transition times are finite, 
and between the transitions the process lives in a small neighbourhoods of wells' minima.   
As opposed to the Gaussian simulated annealing, the L\'evy flights process does not settle
down near the global maximum of $U$. However, our results can be applied 
for a search for the global minimum of the 
potentials, which possess the so-called
``large-rims-have-deep-wells'' property, see \cite{Schoen-97,Locatelli-02}, i.e.\ when the 
spatially largest well is at the same time the deepest. 
Then, having empirical estimates of the local minima locations $m_i$ and the invariant distribution $\pi$, we
can derive the coordinates of the saddle points $s_i$, and thus reconstruct the sizes of the wells.

On the other hand, if the cooling is fast, i.e.\ $\theta>1/\alpha$, the L\'evy particle
gets trapped in one of the wells when the temperature decreases below some critical level.

In this paper we do not answer the most interesting question: 
is it possible to detect a global minimum of $U$ with
help of a non-local search and L\'evy flights? The answer to this question is affirmative, and the results will be presented 
in our forthcoming paper.

\section*{References}

\bibliography{biblio-new}
\bibliographystyle{alpha}

\end{document}